\documentclass[a4]{epl2}
\usepackage{graphicx}
\usepackage{epsfig}
\usepackage{color}
\usepackage{bm}
\usepackage{amssymb}
\usepackage{MnSymbol}
\usepackage{bbm}

\title{\textcolor[rgb]{0.00,0.00,1.00}{{ Dual quantum mechanics and its electromagnetic analog}}}
\shorttitle{\textcolor[rgb]{0.00,0.00,1.00}{ Dual quantum mechanics and its electromagnetic analog }}

\author{ \textcolor[rgb]{0.00,0.00,.00}{A. I. Arbab}\inst{} \footnote{arbab.ibrahim@gmail.com}}

\institute{\inst{}
Department of Physics,
College  of Science, Qassim University, P.O. Box 6644, Buraidah 51452, KSA}

\abstract{ An eigenvalue equation representing symmetric (dual) quantum equation is introduced. The particle is described by two scalar wavefunctions, and two vector wavefunctions. The eigenfunction is found to satisfy the quantum Telegraph equation keeping the form of the particle fixed but decaying its amplitude. An analogy with Maxwellian equations is presented. Massive electromagnetic field will satisfy a quantum Telegraph equation instead of a pure wave equation. This equation resembles the motion of the electromagnetic field in a conducting medium. With a particular setting of the scalar and vector wavefunctions, the dual quantum equations are found to yield the quantized Maxwell's equations. The total energy of the particle is related to the phase velocity ($v_p$) of the wave representing it by $E=p\,|v_p|$\,, where $p$ is the matter wave momentum. A particular solution which describes the process under which the particle undergoes a creation and annihilation is derived. The force acting on the moving particle is expressed in a dual Lorentz-like form. If the particle moves like a fluid, a dissipative (drag) quantum force will arise.}


\begin{document}
\maketitle
\baselineskip=16pt

\section{\textcolor[rgb]{0.00,0.07,1.00}{Introduction}}

The advent of the 20th century had witnessed a great deal of formulations of physical laws. Quantum concepts were formulated into quantum theory. Based on Planck and de Broglie hypotheses, Schr\"{o}dinger equation was born. The electron is demonstrated to have  a spin angular momentum equals to 1/2 \textcolor[rgb]{0.00,0.07,1.00}{\cite{bjorken}}. Based on Einstein relativity, relativistic quantum mechanics were formulated. This is culminated in Dirac and Klein-Gordon equations \textcolor[rgb]{0.00,0.07,1.00}{\cite{bjorken}}. The relativistic electron as described by Dirac, is  designated by four quantum states. It thus represented by a spinor that reflects a state of an electron with spin-up and down states, and a positron with spin-up and down states. In Schr\"{o}dinger paradigm, the electron is described by a single complex wavefunction out of which its dynamical states are found. In this theory, the Born's idea of probabilistic description of the particle is adopted. A theory of particles having a spin of 3/2 is advocated by   Rarita  and Schwinger \textcolor[rgb]{0.00,0.07,1.00}{\cite{rarita}}.

A new formulation that combines Schr\"{o}dinger, Dirac and Klein-Gordon theories is recently presented \textcolor[rgb]{0.00,0.07,1.00}{\cite{epl}}. In such a theory the particle states is represented by a quaternion (generalized complex number). The special nature of quaternion multiplication, that is not commutative, provides a unification between the three paradigms \textcolor[rgb]{0.00,0.07,1.00}{\cite{epl}}. Unlike the Schr\"{o}dinger and Dirac representations,  a particle wavefunction is represented by a 3- vector and a scalar, in the quaternion representation. These are analogous to the Maxwell's scalar and vector potentials. Recall that in Dirac formalism, the electron velocity turns out to be equal to $\pm\,c$, where $c$ is the speed of light in vacuum. However, in biquaternionic formalism it is the group velocity that is equal to $\pm\,c$, and not the velocity of the particle.

We would like here to extend our quaternion wavefunction representation to biquaternion one, where now the scalar and vector parts of the quaternions are complex. In this case  the particle is described by eight components. These components are governed by four equations. This is reminiscent of the Maxwell's representation of the electromagnetic field when magnetic monopoles are included.  We thus provide a description of a quantum particles that is analogous to the Maxwellian description of the classical electromagnetic fields originated from the electric charge. The gauge transformations obeyed by massive field of a quantum particle extends that one that adhered by the Maxwell's equations.

\section{\textcolor[rgb]{0.00,0.07,1.00}{Biquaternionic eigenvalue problem}}

In quantum mechanics a physical observable is obtained from the eigenvalue equation. In the ordinary quantum theory, the operator is Hermitian and the eigenfunction is a complex  scalar wavefunction.  To extend this equation to quaternionic one,  the operator and eigenfunction will become quaternions. Following the spirit of Dirac equation, a symmetric formalism of a moving quantum particle can be described by the biquaternionic eigenvalue of the form \textcolor[rgb]{0.00,0.07,1.00}{\cite{epl}}
\begin{equation}
\tilde{P}\,\,\tilde{\Psi}=mc\tilde{\Psi}\,,\qquad \tilde{P}=\left(\frac{i}{c}\,E\,,\vec{p}\,\right)\,,\qquad\tilde{\Psi}=\left(\frac{i}{c}\,\phi\,,\,\,\vec{\psi}\,\right),
\end{equation}
where $E=i\hbar\frac{\partial}{\partial t}$ and $\vec{p}=-i\hbar\,\vec{\nabla}$ are the total energy (Hamiltonian) and momentum operator of the particle. As in electromagnetism, where the electric charge is described by  scalar and vector potentials, $\varphi$ and $\vec{A}$; we want here to describe the mass (quantum particle) by its two scalar and vector potentials, $\phi$ and $\vec{\psi}$ taking into account their complex nature. In this situation it is more appropriate to employ biquaternions rather than the ordinary quaternions. To this aim, we express them as $\phi=\phi_r+i\phi_t$ and $\vec{\psi}=\vec{\psi}_r+i\vec{\psi}_t$\,. Hence, Eq.(1) becomes
\begin{equation}
\tilde{P}\tilde{\Psi}=mc\tilde{\Psi}\,,\qquad \tilde{P}=(\frac{i}{c}\,E\,,\vec{p})\,,\qquad\tilde{\Psi}=\left(\frac{i}{c}\,\phi_r-\frac{1}{c}\,\phi_t\,,\,\vec{\psi}_r+i\vec{\psi}_t\,\right).
\end{equation}
Expanding Eq.(2) using the quaternion product rule, $\tilde{A}\,\tilde{B}=(a_0b_0-\vec{a}\,\cdot\vec{b}\,, a_0\vec{b}+\vec{a}\,b_0+\vec{a}\times\vec{b}\,)$, for the two quaternions, $\tilde{A}=(a_0\,, \vec{a})$ and $\tilde{B}=(b_0\,, \vec{b})$, we obtain
\begin{equation}
\vec{\nabla}\cdot\vec{\psi}_r-\frac{1}{c^2}\frac{\partial\phi_r}{\partial t}=\frac{m}{\hbar}\,\phi_r\,,
\end{equation}
\begin{equation}
\vec{\nabla}\cdot\vec{\psi}_t-\frac{1}{c^2}\frac{\partial\phi_t}{\partial t}=\frac{m}{\hbar}\,\phi_t\,,
\end{equation}
and
\begin{equation}
\vec{\nabla}\times\vec{\psi}_r+\frac{1}{c}\frac{\partial\vec{\psi}_t}{\partial t}-\frac{1}{c}\,\vec{\nabla}\phi_t=-\frac{mc}{\hbar}\,\vec{\psi}_t\,,
\end{equation}
\begin{equation}
\vec{\nabla}\times\vec{\psi}_t-\frac{1}{c}\frac{\partial\vec{\psi}_r}{\partial t}+\frac{1}{c}\,\vec{\nabla}\phi_r=\frac{mc}{\hbar}\,\vec{\psi}_r\,,
\end{equation}
where we equate the real and imaginary parts in the two sides of the resulting equations to each others. These are coupled differential equations. The set of Eqs.(3) - (6) can be combined in the form
\begin{equation}
\frac{1}{c^2}\frac{\partial^2\psi}{\partial t^2}-\nabla^2\psi+\frac{2m}{\hbar}\,\frac{\partial\psi}{\partial t}+\left(\frac{mc}{\hbar}\right)^2\psi=0\,,
\end{equation}
where $\psi=\phi_r, \phi_t, \vec{\psi}_r,\vec{\psi}_t$\,. Equation (7) can be seen as a dissipative Klein-Gordon equation for spin-0 bosons.  It is a fundamental term expressing the inertia of matter waves that is missed in Schr\"{o}dinger theory. It arises in classical mechanics when a particle moves in a fluid. The fluid here could be the aether that physicists had abandoned long time ago, and reintroduced under the name  vacuum. The third term in the left hand-side of the above equation is the cause of the dissipation in the particle's amplitude.

Substituting a plane wave solution of the form
\begin{equation}
\psi=C \exp i(\vec{k}\cdot\vec{r}-\omega\,t)\,,
\end{equation}
where $C$ is constant, $\vec{k}$ and $\omega$ are the propagation and angular frequency of the wave, respectively, in Eq.(7), one obtains the dispersive relation
\begin{equation}
\omega=-\frac{mc^2i}{\hbar}\pm\, ck\,.
\end{equation}
The imaginary term in Eq.(9) gives rise to dissipation (decay) term in the wavefunction, while the real term will yield an oscillation, \emph{viz}.,
\begin{equation}
\psi=C \exp\,(-mc^2t/\hbar)\,\exp i(\vec{k}\cdot\vec{r}\mp\,ck\,t)\,.
\end{equation}
Hence, our above quantum equation describes a particle by two waves moving to the left and right with velocity of light but with a decaying amplitude owing to the particle mass. Hence, it agrees with the concept of the particle as a wavepacket. If we express the solution of Eq.(7) in the form, $\psi=e^{\alpha\, t}\phi(r,t)$, the $\phi$ will satisfy the Klein-Gordon equation, $\frac{\ddot\phi}{c^2}-\phi''+(\frac{mc}{\hbar})^2\phi=0$, and that $\alpha=-mc^2/\hbar$, therefore, $\psi\propto e^{-mc^2t/\hbar}\phi(r,t)$. Hence, $\psi$ describes a damped harmonic scalar spin-0 field. It is worth mentioning that one can obtain the Dirac equation from Eq.(7) by replacing the mass $m$ by $\pm \,i\,m$ or $r$ by $\pm\,ir$, or $t$ by $\pm\, it$ \textcolor[rgb]{0.00,0.07,1.00}{\cite{epl}}. Now substituting $\psi=e^{\beta r}\xi(r,t)$ in Eq.(7)  yields
$$\frac{1}{c^2}\frac{\partial^2\xi}{\partial t^2}-\nabla^2\xi+\left(\frac{m^2c^2}{\hbar^2}-\beta^2\right)\xi-2\beta\,\frac{\partial\xi}{\partial r}+\frac{2m}{\hbar}\frac{\partial\xi}{\partial t}=0\,.$$
Let us now consider the case $\beta=\pm\,mc/\hbar$,  in the above equation to obtain the wave equation
$$\frac{1}{c^2}\frac{\partial^2\xi}{\partial t^2}-\nabla^2\xi\pm\,\frac{2mc}{\hbar}\,\frac{\partial\xi}{\partial r}+\frac{2m}{\hbar}\frac{\partial\xi}{\partial t}=0\,.$$
Now substitute the plane wave solution, $\xi\sim e^{i(\vec{k}\cdot\vec{r}-\omega\,t)}$,  in the above equation to obtain the dispersion relation
 $$\omega^2+i\frac{2mc^2}{\hbar}\,\omega\mp\,i\frac{2mc^3}{\hbar}\,k-c^2k^2=0\,,\qquad (\omega\pm ck)\left(\omega\mp ck+\frac{2imc^2}{\hbar}\right)=0\,.$$
The solution of the above equations is
$$\omega=\pm\,ck\,,\qquad\qquad \omega=\pm\,ck-i\frac{2mc^2}{\hbar}\,.$$
This leads to two phase  velocities, $v_1=\pm\,c$ and $v_2=\pm\,c-i\frac{2mc^3}{\hbar\,k}$\,, and two group velocities, $v_{g1}=\pm\,c$ and $v_{g2}=\pm\,c$\,. A complex phase velocity (or frequency) of a wave implies that the wave is undergoing attenuation that is proportional to the particle mass. The two phase velocities correspond to a wavepacket composed of two waves; one traveling to the left with velocity of light and a second one with a phase velocity greater than $c$. However, one finds the energies of the two waves representing the particle to be  $E_1=v_1p=cp$ and $E_2=|v_2|p=\sqrt{c^2p^2+(2m)^2c^4}$, where $p=\hbar\,k$, implying that the other wave carries an energy, $E_2=|v_2|p$, having twice the mass of the particle. This is interesting since when a particle and its antiparticle interact (collide) a photon is created. Hence, this particular wave describes the process in which the particle undergoes an annihilation and creation process.  In this case one finds the particle field $\psi=e^{-mcr/\hbar}\,\phi(r,t)$ whose amplitude decays exponentially with distance. Hence, a field, owing to its  matter nature, decays exponentially with distance that may explain how neutrons interact inside a nucleus having a short range while having no electric charge.

The energy conservation associated with Eq.(7) can be found. It is of the form
$$\vec{\nabla}\cdot\vec{S}+\frac{\partial u}{\partial t}=-\frac{2mc^2}{\hbar}\,u\,,\qquad \vec{S}=c^2\hbar\left(\psi\vec{\nabla}\psi^*-\psi^*\vec{\nabla}\psi\right)\,,\qquad u=\hbar\left(\psi^*\frac{\partial\psi}{\partial t}-\psi\frac{\partial\psi^*}{\partial t}\right)\,.$$
The angular frequency for Dirac's particle wave is thus given by
\begin{equation}
\omega_D=\mp\frac{mc^2}{\hbar}\pm\, ck\,,
\end{equation}
which represents  4-wave oscillations. These can be seen as degrees of freedom (states) a particle can have. Transitions between these states could be governed by some physical rule pertaining to energy and momentum conservation. They are internal degrees of freedom, since the group velocity representing the particle is traveling at the speed of light. In the Dirac's original formalism, these correspond to the particle with spin-up and spin-down, and antiparticle with spin-up and spin-down states (the spinors). The 4-oscillations the particle  could lead to spin (rotation), since the projection of a rotational motion is equivalent to  an oscillatory motion.

Now the replacement $m\rightarrow \pm \,i\,m$ in Eq.(7) will yields
\begin{equation}
\frac{1}{c^2}\frac{\partial^2\psi_\pm}{\partial t^2}-\nabla^2\psi_\pm\pm\,\frac{2mi}{\hbar}\,\frac{\partial\psi_\pm}{\partial t}-\left(\frac{mc}{\hbar}\right)^2\psi_\pm=0\,,
\end{equation}
These are the equations for the spin-1/2 particle ($\psi_+$) and antiparticle ($\psi_-$). It is interesting to remark that with the transformation, $m\rightarrow \pm\,i\,m$ (or $r\rightarrow \pm\,i\,r,$ or $ t\rightarrow \pm\,i\,t$) we go from spin-0 particle description to spin-1/2 description. The latter transformation corresponds to rotating the $r-t$ plane by an angle of $\pi/2$ clockwise and counterclockwise. In the Cooper pairs mechanism the two electrons are compelled to behave as a boson and they actually did. This idea was codified by Ginzburg and Landau in 1950, who introduced an order parameter field describing the condensate of electrons \textcolor[rgb]{0.00,0.07,1.00}{\cite{landau}}. Similarly, in Bose condensation, fermions are required to behave as boson to go into the condensation state.  Hence, there appears an urgent need for such a situation to exit. Hence, the above mathematical manipulation helps accomplish that. Therefore, when such a condition is needed, the wave equation of the fermionic state should transform to bosonic state by one of the three mentioned methods. Notice that the Dirac original equation is a first order differential equation. This is new version of Dirac equation bearing the general structure of a wave equation that is of second order form. The two solutions, $\psi_\pm$ are oscillating waves and no dissipation term exists. This why a quantum particle wave due to its mass alone has a short range character, while Dirac's particle wave has an infinite range. It is remarkable that the Schrodinger equation is a late time aspect of Eq.(12) where the first term becomes negligible. In this case Eq.(12) reduce to the form
$$
\mp i\hbar\,\frac{\partial\psi_\pm}{\partial t}=-\frac{\hbar^2}{2m}\,\nabla^2\psi_\pm-\frac{mc^2}{2}\psi_\pm\,,
$$
which is the Schrodinger equation representing a particle with positive and negative non-relativistic energy states, and with a potential energy equals to $V=-mc^2/2$ for the two energy states. This negative potential energy could imply that a particle is impeded  by its unsurmountable mass (inertia). Note that the ordinary Schrodinger equation represents only a particle with positive energy. Hence, the above equation can be seen as an extended  form of Schrodinger equation governing a free quantum particle and its antiparticle at late times of the particle evolution.

The solution in Eq.(10) resembles that one due to the propagation of the electric field in a conducting medium, where \textcolor[rgb]{0.00,0.07,1.00}{\cite{analog}}
\begin{equation}
\sigma=\frac{2m}{\mu_0\hbar}\,,
\end{equation}
where $\sigma$ and $\mu_0$ are the electric conductivity and permeability of the medium. Hence, the effect of the electric conductivity to the electromagnetic system is tantamount to the mass of a quantum system. Equation (13) can be inverted to give the mass of the electromagnetic field (photon) inside a conducting medium. In terms of the electron mass, $m_e$, the photon mass will be, $m=(\frac{\mu_0\hbar\,\sigma}{2m_e})\,m_e$\,. An experimental verification can thus be set.  Note that the skin depth for a conductor when irradiated with light of frequency $\omega$ is $\delta=\sqrt{2/\mu_0\sigma\omega}$. Hence, the photon (light) mass inside a conductor can be calculated as, $m=\hbar/(\delta^2\omega)$.

Using Eqs.(14) and (15), we see that the particle velocity, $v$, phase velocity, and group velocity are related by the relation
$$v\,|v_p|v_g=c^3\,.$$
The group and phase velocities of the particle represented by Eq.(7) are, respectively
\begin{equation}
v_p=\frac{\omega}{k}=-\frac{mc^2i}{\hbar\,k}\pm c\,,\qquad\qquad v_g=\frac{d \omega}{d k}=\pm\, c\,.
\end{equation}
Now Eq.(14) yields
$$p\, |v_p|=\sqrt{m^2c^4+c^2p^2}=E\,,\qquad p=\hbar\,k\,.$$
It interesting that the total energy of the particle, $E=p\,|v_p|$, is related to its phase velocity as  light energy related to its speed, $E=p\,c$. Note that Eq.(14) states that when $m=0$ then $v_p=c$.

The phase velocity can be expressed as
\begin{equation}
v_p=-i\left(\frac{\lambda}{\lambda_c}\right)c\pm c\,,\qquad \lambda=\frac{2\pi}{k}\,,\qquad \lambda_c=\frac{h}{mc}\,.
\end{equation}
The complex nature of the phase velocity reflects the wave-particle duality of the quantum particle, where the imaginary term has to do with the matter nature, while the real term is related to the wave pure nature. At any moment the particle bears a dual nature. This is what de Broglie had found in 1920. For $\lambda>>\lambda_c,$ the matter component of the particle is more than the wave component.  However, a particle with $\lambda<<\lambda_c$ can be seen as a pure wave.

Equation (7) is analogous to the Telegraph equation governing the propagation of electric signals in transmission lines. It describes waves traveling to the left and right with speed $c$ without distortion, but with a decaying amplitude. Equation (2) generalizes our earlier quantum wave equation \textcolor[rgb]{0.00,0.07,1.00}{\cite{epl}}. Note that the electric signals waves are  response of the electrons (due to their charge nature) in the wires. Hence, the motion of electrons due to their charge nature is now compatible (coherent) due to its mass (quantum) nature. Therefore, Eq.(4) unifies the electromagnetic and quantum properties of moving particles.  While the transmission equation presumes the charge nature the electrons, the quantum analogue does'nt.

\section{\textcolor[rgb]{0.00,0.07,1.00}{Maxwell electrodynamics and quantum analogy}}

It is interesting to see that the matter potentials equations, Eqs.(3) - (6), are analogous to symmetric Maxwell's equations that are obtained when the magnetic monopoles are presumed to exist \textcolor[rgb]{0.00,0.07,1.00}{\cite{monop,monop1}}. Therefore, matter waves allow material monopoles to exist a priori. Moreover, Eqs.(3) - (6) allow a symmetry (material duality) in which
\begin{equation}
\phi_r\rightarrow \phi_t\,,\qquad \vec{\psi}_r\rightarrow \vec{\psi}_t\,,\qquad \phi_t \rightarrow -\phi_r\,,\qquad \vec{\psi}_t\rightarrow -\vec{\psi}_r\,.
\end{equation}
Because of the above dual symmetry, we refer to Eqs.(3) - (6) as dual quantum equations. They generalize our earlier equations of quantum mechanics \textcolor[rgb]{0.00,0.07,1.00}{\cite{epl}}.
A gauge-like transformation of the form \textcolor[rgb]{0.00,0.07,1.00}{\cite{qte}}
\begin{equation}
\vec{\psi}_r\,'=\vec{\psi}_r-\vec{\nabla}\Lambda\,,\qquad\qquad \phi_r'=\phi_r-\frac{\partial\Lambda}{\partial t}-\frac{mc^2}{\hbar}\,\Lambda\,,
\end{equation}
and
\begin{equation}
\vec{\psi}_t\,'=\vec{\psi}_t+\vec{\nabla}\lambda\,,\qquad\qquad \phi_t'=\phi_t+\frac{\partial\lambda}{\partial t}+\frac{mc^2}{\hbar}\,\lambda\,,
\end{equation}
makes Eqs.(3) - (6) intact, where $\Lambda$ and $\lambda$ satisfy  Eq.(7). We better call this the \emph{massive gauge transformation}.

One can now compare Eq.(3) with the massive Lorenz gauge condition if we set
\begin{equation}
\vec{\psi}_r\rightarrow\vec{A}\,,\qquad\qquad \phi_r\rightarrow -\varphi\,,
\end{equation}
where we obtain  \textcolor[rgb]{0.00,0.07,1.00}{\cite{analog}}
\begin{equation}
\vec{\nabla}\cdot\vec{A}+\frac{1}{c^2}\frac{\partial\varphi}{\partial t}=-\frac{m}{\hbar}\,\varphi\,.
\end{equation}
Therefore, Eqs.(3) and (4) can be seen as the Lorenz gauge conditions for the matter fields, which reduce to the electromagnetic Lorenz gauge condition when $m=0$.

Now apply the replacement in (19), in Eqs.(5) and (6) to obtain a matter field that is analogous to the electric field, $\vec{E}_m$ defined as
\begin{equation}
\vec{E}_m=-c\vec{\nabla}\times\vec{\psi}_t\,,\qquad\qquad \vec{B}_m=\vec{\nabla}\times\vec{\psi}_r\,,
\end{equation}
where
\begin{equation}
\vec{E}_m=\vec{E}-\frac{mc^2}{\hbar}\, \vec{A}\,,\qquad\vec{E}=-\vec{\nabla}\varphi-\frac{\partial\vec{A}}{\partial t}\,,\qquad \vec{B}_m=\vec{\nabla}\times\vec{A}=\vec{B}\,,
\end{equation}
and
that reduces to the ordinary electric field when $m=0$. It is time now to seek the physical significance of the quantities that describe the quantum particle, \emph{viz.},  $\phi_r, \phi_t, \vec{\psi}_r,\vec{\psi}_t$\,.

Now taking the curl of Eq.(5) and (6) yields
\begin{equation}
\vec{\nabla}\times\vec{B}_m=\frac{1}{c^2}\,\frac{\partial\vec{E}_m}{\partial t}+\frac{m}{\hbar}\,\vec{E}_m\,,
\end{equation}
and
\begin{equation}
\vec{\nabla}\times\vec{E}_m=-\frac{\partial\vec{B}_m}{\partial t}-\frac{mc^2}{\hbar}\,\vec{B}_m\,.
\end{equation}
Now taking the divergence of Eq.(21) yields
\begin{equation}
\vec{\nabla}\cdot\vec{E}_m=0\,,\qquad\qquad \vec{\nabla}\cdot\vec{B}_m=0\,,
\end{equation}
are the material Gauss's laws for electricity and magnetism, where $\vec{E}_m$ and $\vec{B}_m$ satisfy Eq.(7). They imply that  the electric and magnetic matter fields are divergenceless. Equation (7) suggests that the transformation $\frac{\partial}{\partial t}\rightarrow \frac{\partial}{\partial t}+\frac{mc^2}{\hbar}$ in a wave equation provides a mass for a massless scalar field under consideration. For fermionic field one has the transformation, $\frac{\partial}{\partial t}\rightarrow \frac{\partial}{\partial t}\pm\frac{imc^2}{\hbar}$.  It can be used to provide mass to massless boson (spin-0) and fermion (spin-1/2) fields. This is evident from Eq.(23) and (24). One very important point pertaining to the quantum-Maxwell correspondence is that the Maxwell's scalar and vector potentials satisfy two different wave equations (except when the Lorenz gauge condition is considered) in comparison with the quantum potentials that are both  governed by Eq.(7). If we drop the subscript `m' in Eqs.(23) - (25), we obtain the quantized Maxwell's equations that we recently derived \textcolor[rgb]{0.00,0.07,1.00}{\cite{quantized}}. When $m=0$\,, we recover the ordinary Maxwell's equations.

Comparing Eqs.(23) and (24) with Ampere and Faraday equations for electromagnetism would imply two currents are associated with the electric and magnetic fields as
\begin{equation}
\vec{J}_E=\sigma_m\,\vec{E}_m\,,\qquad \qquad \vec{J}_B=\frac{mc^2}{\hbar}\,\vec{B}_m\,,\qquad \sigma_m=\frac{m}{\mu_0\hbar}\,.
\end{equation}
Equations (23) - (24) are similar to Maxwell's equations for an electromagnetic field inside a non-charged conductor.
These equations are recently found in the framework of quantized Maxwell's equations \textcolor[rgb]{0.00,0.07,1.00}{\cite{quantized}}. It is interesting to see that the inclusion of the particle mass doesn't violate the gauge invariance, as evident from  Eq.(23) and (24). These equations show that if the photon is massive, then a current $\vec{J}_m$ will be immediately developed in the system. Hence, a transverse electric field could give rise to a traverse potential as in the Hall effect \textcolor[rgb]{0.00,0.07,1.00}{\cite{hall}}. It is very interesting that the dual quantum mechanics can be put in the forms of that of Maxwell's equations. Thus, our dual representation of quantum mechanics provides a one-to-one correspondence between quantum mechanics and electrodynamics.

Let us consider the situation if we define, $\vec{\psi}_r=\vec{E}_m$ and $\vec{\psi}_t=c\vec{B}_m$\,, and substitute these in Eqs.(3) -(6) to obtain
\begin{equation}
\vec{\nabla}\cdot\vec{E}_m=\frac{1}{c}\,\frac{\partial\phi_r}{\partial t}+\frac{mc}{\hbar}\,\phi_r\,,
\end{equation}
\begin{equation}
\vec{\nabla}\cdot\vec{B}_m=\frac{1}{c^2}\frac{\partial\phi_t}{\partial t}+\frac{m}{\hbar}\,\phi_t\,,
\end{equation}
and
\begin{equation}
\vec{\nabla}\times\vec{E}_m=-\frac{\partial\vec{B}_t}{\partial t}+\vec{\nabla}\phi_t-\frac{mc^2}{\hbar}\,\vec{B}_m\,,
\end{equation}
\begin{equation}
\vec{\nabla}\times\vec{B}_m=\frac{1}{c^2}\frac{\partial\vec{E}_m}{\partial t}-\frac{1}{c}\,\vec{\nabla}\phi_r+\frac{m}{\hbar}\,\vec{E}_m\,.
\end{equation}
Here $\vec{J}_t=-\vec{\nabla}\phi_t$ and $\vec{J}_r=-\frac{1}{\mu_0c}\vec{\nabla}\phi_r$ can be seen as magnetic and  electric diffusion current densities, respectively. It  thus becomes apparent that the $\phi_r$ and $\phi_t$ in our particle representations are responsible for diffusion currents. They can be related to the electric and magnetic charge densities as $$\phi_r=\frac{\hbar}{mc}\,\rho_e\,,\qquad\qquad \phi_t=\frac{\hbar}{m}\,\rho_m\,.$$
Hence, any spatial and temporal variations of these scalars will result immediately in inducing charge and current densities. Equations (27) - (30) are also found in the framework of quantizing Maxwell's equations \textcolor[rgb]{0.00,0.07,1.00}{\cite{quantized}}. Consequently once can assume that the massive electromagnetic field is governed by same equations except that $\vec{E}_m\rightarrow \vec{E}$ and $\vec{B}_m\rightarrow \vec{B}$\,. The resulting equations will describe massive dual electrodynamics, where $\phi_r\rightarrow \varphi$, and $\phi_t\rightarrow \psi$, are the electric and magnetic potentials in symmetric Maxwell electrodynamics. It is interesting that magnetic monopoles arise naturally in formulating matter waves. This is unlike the standard formalism of Maxwell's equations where magnetic monopoles are assumed not to exist.

Because of the last term on the right hand-side of Eq.(29), a static magnetic field can lead to an electromotive force given by $\epsilon_e=\oint \vec{E}\cdot d\vec{\ell}=\frac{mc^2}{\hbar}\, \Phi_B$, where $\Phi_B$ is the magnetic field. The Hall voltage across slab of thickness $d$ when a normal magnetic field is applied on it is given by $V_H=IB/(nqd)$, where $n$ is number of charges per unit volume. Now one finds, $V_H=\Phi_BI/(qndA)$, and hence, for a  flux quantum of $\Phi_0=\hbar/q$, we obtain $V_H=I\hbar/Nq$, where $N$ is the number of particles present. Thus, if $qV_H=mc^2$, then a mass quantum associated with the current $I$ will be
$$m=\frac{\hbar\,I}{c^2Nq}\,.$$
This formula can also be obtained if we set $mc^2=qV$, in quantum Hall effect, where $V=IR$, for a resistance quantum   given by $R=\hbar/q^2$.  In this connection, we derived an analogous formula for the mass  of the photon created by a gravitational field of the black hole, $g$, given by $m=\frac{\hbar}{c^2}\,\frac{g}{2c}$ \textcolor[rgb]{0.00,0.07,1.00}{\cite{blackhole}}.
Such an analogy implies that the gravitational acceleration (field) is $g=cI/q$, for $N=2$. This agrees with a current that is found to be associated with the Cooper pairs (massive photo) motion in Josephson junction \textcolor[rgb]{0.00,0.07,1.00}{\cite{josephson0,current-jose}}. In the standard picture, the supercurrent is assumed to
result from the coherent tunneling of Cooper pairs driven by the phase difference  between the  the two superconductors. The above equation can be deduced using the Heisenberg's uncertainty relation, $\Delta E\Delta\, t\ge\hbar/2$, with $\Delta E=mc^2$ and $\Delta t=q/I$, so that one obtains, $mc^2q/I=\hbar/2$ or $m=\hbar I/2c^2q$. However, Josephson energy is but given by $E_J=\hbar I/2q=mc^2$, where $I$ is known as the critical current \textcolor[rgb]{0.00,0.07,1.00}{\cite{josephson0}}. We assume the quantum effect to be due to the propagation of a massive photon that gives rise to a quantum current flow inside the Josephson junction.

Does the analogy between our quantum particle description and the Maxwellian description allow use to describe the massive electromagnetic fields by Eqs.(3) - (6) instead of the Maxwell- Proca equations that break gauge invariance? If so, then the electromagnetic field will satisfy the equations  \textcolor[rgb]{0.00,0.07,1.00}{\cite{analog}}
\begin{equation}
\frac{1}{c^2}\frac{\partial^2\vec{E}}{\partial t^2}-\nabla^2\vec{E}+\frac{2m}{\hbar}\,\frac{\partial\vec{E}}{\partial t}+\left(\frac{mc}{\hbar}\right)^2\vec{E}=0\,,
\end{equation}
and
\begin{equation}
\frac{1}{c^2}\frac{\partial^2\vec{B}}{\partial t^2}-\nabla^2\vec{B}+\frac{2m}{\hbar}\,\frac{\partial\vec{B}}{\partial t}+\left(\frac{mc}{\hbar}\right)^2\vec{B}=0\,.
\end{equation}

\section{\textcolor[rgb]{0.00,0.07,1.00}{The quantum energy  conservation equation}}

The energy  conservation equation of the quantum system, Eqs.(3) - (6), can be obtained  by multiplying (dotting)  Eq.(5) by $\vec{\psi}_t$ and Eq.(6) by $\vec{\psi}_r$, and subtracting the two resulting equations to obtain
\begin{equation}
\frac{\partial u}{\partial t}+\vec{\nabla}\cdot\vec{S}=-\frac{mc^2}{\hbar}\,u\,,
\end{equation}
where
\begin{equation}
u=c^2\psi^2_r+c^2\psi^2_t+\phi_r^2+\phi^2_t\,,\qquad \vec{S}=c^3\vec{\psi}_r\times\vec{\psi}_t-c^2\phi_r\vec{\psi}_r-c^2\phi_t\vec{\psi}_t\,.
\end{equation}
It is thus very interesting to remark that all the scalar and vector potentials representing the quantum particle contribute to its energy. The particle energy flows along the $\vec{\psi}_r$ and $\vec{\psi}_t$ directions, and a direction normal to them. The energy equation in Eq.(33) is invariant under the duality transformation set in Eq.(16). The dissipation of particle energy,  the right hand-side of Eq.(33), is due to the particle mass. Therefore, a massless particle wave will propagate without dissipation. This can be seen as if the particle moves in a viscous fluid that is being affected by the drag force. In electrodynamics, an accelerating $(a$) charge radiates with radiation power given by Larmor formula, $P_L=\frac{\mu_0qa^2}{6\pi c}$. So one would expect to have an analogous radiation to be connected with an accelerating neutral particle whose power is given by $P_m=\frac{\hbar a^2}{2c^2}$ \textcolor[rgb]{0.00,0.07,1.00}{\cite{blackhole}}.
It is worth to mention that in electromagnetism, we have the field equations given by the Maxwell's equations. The force equation on the particle is given by the Lorentz force.  With the same token, we derived here the field equations, Eqs.(3) - (6) and we want now to derive the force equation acting on the particle mass. Let us call this the  quantum equation of motion. It is the analogue of Lorentz force for a quantum particle expressed in terms of two fields. Owing to the quantum - electromagnetism analogy outlined above, one is advised to express it in the following form \textcolor[rgb]{0.00,0.07,1.00}{\cite{essay}}
\begin{equation}
\tilde{F}=-m\tilde{V}\tilde{\nabla}\tilde{\Psi}\,,\qquad \tilde{V}=(ic\,, \vec{v}\,)\,,\qquad \tilde{\Psi}=\left(\frac{i}{c}\,\phi_r-\frac{1}{c}\,\phi_t\,,\,\, \vec{\psi}_r+i\,\vec{\psi}_t\,\right),
\end{equation}
where $\vec{v}$ is the velocity of the quantum particle. In this case, $\vec{\psi}_r$ \& $\vec{\psi}_t$ act like velocity, and $\phi_r$ \& $\phi_t$ like potential with dimension of $[c^2]$. They are like $\vec{A}$ and $\varphi$ in electromagnetism, where $\vec{A}$ were assumed to represent  some kind of fluid velocity field.

The force equation associated with Eqs.(27) - (30) can be expressed as
\begin{equation}
\tilde{F}=-\frac{m}{c}\tilde{V}\tilde{\Psi}\,,\qquad \tilde{V}=(ic\,, \vec{v}\,)\,,\qquad \tilde{\Psi}=\left(\frac{i}{c}\,\phi_r-\frac{1}{c}\,\phi_t\,,\,\, \vec{\psi}_r+i\,\vec{\psi}_t\,\right),
\end{equation}
which upon using the quaternionic expansion, yields the two forces
\begin{equation}
\vec{f}_r=-m(\vec{\psi}_t+\frac{\vec{v}}{c^2}\,\,\phi_t)+m\frac{\vec{v}}{c}\times\vec{\psi}_r\,,\qquad\qquad \vec{f}_t=m(\frac{\vec{v}}{c^2}\,\,\phi_r+\vec{\psi}_r)+m\frac{\vec{v}}{c}\times\vec{\psi}_t\,,
\end{equation}
and the two powers
\begin{equation}
P_r=-mc(\phi_t+\vec{v}\cdot\vec{\psi}_t)\,,\qquad\qquad P_t=mc(\phi_r+\vec{v}\cdot\vec{\psi}_r)\,,
\end{equation}
where
\begin{equation}
\tilde{F}=(\frac{i}{c}P_r-\frac{P_t}{c}\,\,, \vec{f}_r+i\,\vec{f_t}).
\end{equation}
In terms of the matter fields, $\vec{\psi}_r=\vec{E}_m$ and $\vec{\psi}_t=c\vec{B}_m$, Eqs.(37) and (38) yield
\begin{equation}
\vec{f}_r=mc(\vec{B}_m-\frac{\vec{v}}{c^2}\times\vec{E}_m)+\frac{m\vec{v}}{c^2}\,\,\phi_t\,,\qquad\qquad \vec{f}_t=m\left(\vec{E}_m+\vec{v}\times\vec{B}_m\right)-\frac{m\phi_r}{c^2}\,\vec{v}\,\,,
\end{equation}
and
\begin{equation}
P_r=m(\phi_t-c\vec{v}\cdot\vec{B}_m)\,,\qquad\qquad P_t=m(-\phi_r+\vec{v}\cdot\vec{E}_m)\,.
\end{equation}
Now Eq.(40) is a Lorentz-like force acting on the moving quantum particle. This is why we defined the quaternionic force in Eq.(36). The last term in the right hand-side of the matter Lorentz force, $\vec{f}_t$ is a viscous (drag) force due to the inertia (mass) of the particle. It is interesting to remark that Eqs.(37) and (38) are invariant under the matter duality transformation in Eq.(16), where $\vec{f}_r\rightarrow \vec{f}_t$ and $P_r\rightarrow P_t$ (and $\vec{f}_t\rightarrow -\vec{f}_r$ and $P_t\rightarrow -P_r$).

Recall that the force on a moving mass (fluid) can be described by $\tilde{F}=-m\,\tilde{V}\tilde{\nabla}\,\tilde{V}$\textcolor[rgb]{0.00,0.07,1.00}{\cite{essay}}, and consequently for a quantum particle one suggests  $\tilde{F}=m\,\tilde{V}\tilde{\nabla}^*\,\tilde{\Psi}$. This latter force yields  the two quantum forces
\begin{equation}
\vec{f}_r=-\frac{m^2c}{\hbar}(c\vec{\psi}_r+\vec{v}\times\vec{\psi}_t)\,,\qquad\qquad \vec{f}_t=\frac{m^2c}{\hbar}\,(-c\vec{\psi}_t+\vec{v}\times\vec{\psi}_r)\,,
\end{equation}
and the two powers
\begin{equation}
P_r=\frac{m^2c^2}{\hbar}\,(\phi_r-\vec{v}\cdot\vec{\psi}_r)\,\qquad\qquad P_t=\frac{m^2c^2}{\hbar}\,(\phi_t-\vec{v}\cdot\vec{\psi}_t)\,.
\end{equation}
It is now intriguing to compare the forces in Eq.(40) with those in Eqs.(42). The force representation in Eq.(36) leads to a drag force acting on the moving particle.
If for a special case one chooses, $\tilde{\Psi}=\tilde{V}=(ic\,, \vec{v})$, where $\vec{v}_t=0\,,\phi_t=0\,, \phi_r=c^2\,, \vec{\psi}_r=\vec{v}$ and $\vec{\psi}_t=0$\,, one finds
\begin{equation}
\vec{f}_r=-\frac{m^2c^2}{\hbar}\,\vec{v}\,,\qquad\qquad \vec{f}_t=0\,,\qquad P_r=\frac{m^2c^4}{\hbar}\, \left(1-\frac{v^2}{c^2}\right)\,.
\end{equation}
It is interesting to see that a freely moving quantum particle experiences  a dissipative (drag) force. This is because of the particle mass which acts as if the particle moves in a fluid. Such a fluid was once proposed by physicists but later on rejected. It is felt by massive objects only, and not light as demonstrated by the Michelson-Morley experiment. Unlike the Newtonian force that is proportional to the particle mass, this quantum force is proportional to the square of the particle mass. Recall that no force acts on a freely moving particle in the Newtonian mechanics.
The above quantum force can be understood if one assume the particle mass is not constant. In this case the force acting on a quantum particle can be expressed as $\vec{f}=m\frac{d\vec{v}}{dt}+\vec{v}\,\frac{dm}{dt}$\,. Thus, a freely moving particle will be described by $\vec{f}_q=\vec{v}\,\frac{dm}{dt}$\,. However, since a quantum particle undergoes a process of perpetual state of creation and annihilation, during a time governed by the uncertainty principle, $\Delta E\,\Delta t=\hbar$, then $\frac{dm}{dt}=-m/\tau$, where $\tau=\Delta\, t$ and $\Delta E=mc^2$ so that $\tau=\hbar/mc^2\,$. Then one can write $\vec{f}_q=-\frac{m^2c^2}{\hbar}\,\vec{v}$, as derived above. If one assumes this force to be due to the gravitational force, then one can write $Gm^2/R^2=m^2c^2v/\hbar$, where $R$ is the particle radius. Hence, the gravitational radius of the particle will be, $R=(G\hbar/vc^2)^{1/2}$, that is independent  of the particle mass. At a limiting speed, $v=c$, and the radius $R$ becomes equal to the Planck's length (radius).

One can now express the force acting on a quantum particle can be expressed as
\begin{equation}
\vec{f}=m\frac{d\vec{v}}{dt}-\frac{m^2c^2}{\hbar}\,\vec{v}\,.
\end{equation}
The second term in Eq.(45) can be assumed to be connected with the particle self-interaction.
It is noted by Schrodinger that the electron in Dirac's theory executes a jitter motion though it is utterly free. The cause of such motion can now be connected with the above quantum force.  The maximum force that can act on a freely moving mass is $f_{qMax.}=m^2c^3/\hbar$\,. This force is thought to arise when the quark-gluon plasma is formed. This is the same force as that acted on a charge due to Schwinger's electric field \textcolor[rgb]{0.00,0.07,1.00}{\cite{rarita,schwinger}}. It is thus apparent that the inertial quantum force is equal to the Schwinger electric force. Hence, a particle acted by a force beyond this force may no longer remain stable and is likely to disintegrate. Would we say that this force is that force which holds the proton or the electron?

Let us now  define two fields by $\vec{E}_d=-\frac{mc^2}{\hbar}\,\vec{\psi}_r$ and $\vec{B}_d=-\frac{mc}{\hbar}\,\vec{\psi}_t$\, so that Eq.(42) becomes
\begin{equation}
\vec{f}_r=m(\vec{E}_d+\vec{v}\times\vec{B}_d)\,,\qquad\qquad \vec{f}_t=mc\,(\vec{B}_d-\frac{\vec{v}}{c^2}\,\times\vec{E}_d)\,.
\end{equation}
This force is analogous to the Lorenz force acting on moving electric and magnetic charges.
The two powers can be expressed as
\begin{equation}
P_r=m\vec{v}\cdot\vec{E}_d+\frac{m^2c^2}{\hbar}\,\phi_r\,,\qquad\qquad P_t=mc\vec{v}\cdot\vec{B}_d+\frac{m^2c^2}{\hbar}\,\phi_t\,.
\end{equation}
The power above is no longer given by $P=\vec{f}\cdot\vec{v}$, as usually the case.

\section{\textcolor[rgb]{0.00,0.07,1.00}{Concluding remarks}}

We extended in this work our earlier quaternion quantum mechanics to a biquaternionic one. In this case the particle wave function is described by two vector wavefunctions (potentials), and two scalar potentials. The two vector wavefunctions are analogous to the electric and magnetic fields in Maxwell's formalism. The two scalars lead to diffusion currents associated with the particle motion. The vector and scalar wavefunctions are found to satisfy a Telegraph-like equations, instead of the Schr\"{o}dinger equation. It is a similar equation that electric signals, caused by the electron movement,  follow in the transmission lines. This formalism yields a symmetric (dual) quantum mechanics that encompasses the Schr\"{o}dinger, Dirac, and Klein-Gordon equations. It is analogous to the symmetric Maxwell's equations when magnetic monopoles are incorporated. The dual quantum mechanics is found to be invariant under massive gauge transformations. The wavefunctions describing the dual quantum mechanics are analogous to the scalar and vector potentials of the dual electromagnetic theory. The effect of particle mass in the former is tantamount to the electric conductivity in the latter. The dual quantum  equations are found to reduce to that of the quantized Maxwell's equations we recently derived. The energy of the quantum particle is related to its phase velocity as light energy is related to its speed, \emph{viz}., $E=p\,|v_p|$. A transformation of the dissipative Klein-Gordon equation into Dirac equation is found where spin-1/2 particle and antiparticles are rotation of the bosonic field by an angle of $\pi/2$. A particular solution which describes the process under which the particle undergoes a creation and annihilation is derived. In this case the particle wave decays exponentially with distance. A representation in which the force acting on the particle is that of Lorentz force is obtained. If the motion of the quantum particle is like that of a fluid, a dissipative quantum force arises. The quantum force acting on the moving particle is analogues to the Lorentz force acting on moving electric and magnetic charges.

\section{\textcolor[rgb]{0.00,0.07,1.00}{Acknowledgements}}
I would like to thank Dr. I. Tomsah for reading the manuscript and  enlightening discussion.

\end{document}